\documentclass
[preprint,byrevtex,nofootinbib,10pt,twocolumn,balancelastpage,titlepage,nobibnotes]{revtex4}%
\usepackage{amsfonts}
\usepackage{amsmath}
\usepackage{amssymb}
\usepackage{graphicx}%
\providecommand{\U}[1]{\protect\rule{.1in}{.1in}}
\ifx\pdfoutput\relax\let\pdfoutput=\undefined\fi
\newcount\msipdfoutput
\ifx\pdfoutput\undefined\else
\ifcase\pdfoutput\else
\ifx\paperwidth\undefined\else
\ifdim\paperheight=0pt\relax\else\pdfpageheight\paperheight\fi
\ifdim\paperwidth=0pt\relax\else\pdfpagewidth\paperwidth\fi
\fi\fi\fi
\begin{document}
\preprint{ }
\preprint{UATP/2107}
\title{Maxwell's Demon must remain sebservient to Clausius's statement}
\author{P.D. Gujrati}
\affiliation{$^{1}$Department of Physics, $^{2}$Department of Polymer Science, The
University of Akron, Akron, OH 44325}
\email{pdg@uakron.edu}

\begin{abstract}
Using classical thermodynamics, we argue that Maxwell's demon loses its battle
against Clausius as any temperature difference or other thermodynamic forces
it creates is immediately compensated by spontaneous counterbalancing flows
that bring about equilibration by slower particles in principle. Being
constrained by these spontaneously generated equilibration processes in which
he actively but unwittingly participates, the demon is incapable of destroying
equilibrium and violating the second law. In fact, our investigation shows
that he is unintentionally designed to support it, and does not alter the temperature.

\end{abstract}
\date{\today}
\maketitle

Maxwell's demon, which had been puzzling scientists since 1867 when Maxwell
proposed it in a letter to Tait \cite{Knott}, stands between two neighboring
chambers $\Sigma_{1}$ and $\Sigma_{2}$ (having fixed and identical volumes)
sharing a wall and forming an isolated system $\Sigma$ initially in
equilibrium (EQ) \cite{Maxwell} at a temperature $T_{0}$; see also
\cite{Smoluchowski,Feynman}. The wall has a small hole that the demon D can
open or close at will to select faster particles to go from $\Sigma_{2}$\ into
$\Sigma_{1}$ and slower particles from $\Sigma_{1}$\ into $\Sigma_{2}$.
Maxwell conjectured that the demon, with this ability, raises the temperature
of $\Sigma_{1}$ over $\Sigma_{2}$ without any expenditure of work, which
violates the second law \cite{Maxwell}. Because of this mortal threat to the
basic foundation of classical thermodynamics, the demon has generated a
tremendous amount of debate and some confusion among the best minds of our
time since its inception \cite{Rex}, and has been a constant source of major
conceptual advances and some challenges in theoretical physics
\cite{Brillouin,Landauer,Szilard,Bennett,Plenio} and in the philosophy of
science \cite{Norton,Bennett0}. The main source of confusion has been the
concept of any work done by the demon, and has required the concept of
information entropy, Landauer's principle, minimum dissipation, etc. The demon
may be seeing a resurgence as the role of fluctuations has become more
prominent. Therefore, it is necessary to revisit the demon in light of the
recent understanding of nonequilibrium thermodynamics \cite{Gujrati-Review} as
the concept of work and of the second law in extended state space necessary
for nonequilibrium (NEQ) states have become clarified only recently
\cite{Gujrati-Heat-Work}.

The observable $\mathbf{X}\doteq\mathbf{(}N,E\mathbf{)}$ of $\Sigma$ forms the
state space $\mathfrak{S}_{\mathbf{X}}$ in which EQ states $\mathcal{M}%
_{\text{eq}}$ reside; here $N$ is the single species particle number (so no
chemical reaction) and $E$ is the energy \cite{Callen,Landau}. We do not
consider the volume as it is kept fixed here. The NEQ states $\mathcal{M}$
reside in an extended state space $\mathfrak{S}_{\mathbf{Z}}$, augmented by
\emph{internal variables} \cite{Kestin,deGroot,Maugin,Prigogine} shown
collectively by $\boldsymbol{\xi}$ ; here, $\mathbf{Z}=\{\mathbf{X,}%
\boldsymbol{\xi}\}$ is the NEQ state variable. We assume $\Sigma_{1}$ and
$\Sigma_{2}$ to be quasi-independent so that their entropies are additive.

Simply put, our strategy will be the following. We treat $\Sigma_{1}$ and
$\Sigma_{2}$, the wall, the hole in it, and the demon as the isolated system
$\Sigma$ considered by Maxwell, for which $dE=0$ so the first law is not
useful. Only the second law will play a role. We will assume that the wall,
hole, and D have no interesting thermodynamics just as Maxwell had considered.
This is similar to treating the piston in a cylinder containing a fluid as
having no relevant thermodynamics \cite{Prigogine,Landau}. The demon D
attempts a spatial inhomogeneity in $E$ and $N$ in $\mathcal{M}_{\text{eq}}$,
which results in a NEQ state $\mathcal{M}$ and reduces its entropy. From
Postulate II of Callen \cite{Callen}, flows of energy and particles in both
directions in a wall permeable to them always balance out in $\mathcal{M}%
_{\text{eq}}$, ensure the maximum possible entropy of $\mathcal{M}_{\text{eq}%
}$, and allow fluctuations in them in each chamber. Maxwell does not specify
the nature of the wall, while most workers take it to be impervious to the
flow of energy and particles. This case is covered by the above wall by making
it impervious to both flows. The open hole always results in permeation to
both flows so this case is no different from when the wall is permeable. We
should remark that as soon as D opens the hole for a certain particle velocity
\cite{Maxwell}, he may not even have any mechanical control over all other
particles that will pass through the open hole in both directions with no
restriction on their energies. Nevertheless, we will also consider the case
where this additional flow is forbidden. Indeed, we pay close attention to the
last case, not only because it is what is usually considered, but also because
it reveals some surprising facts about the demon and clarifies its role,
hitherto unknown. The most important aspect of our strategy is the use of the
NEQ thermodynamic temperature that, as we show, explicitly satisfies the
Clausius statement and its extension. This ensures that the entropy generation
brings the system back to EQ but this fact has never been discussed for the
demon paradox.

For simplicity, we will use the term "body" and denoted by $\Sigma_{\text{b}}$
to refer to any one of $\Sigma,\Sigma_{1}$, and $\Sigma_{2}$ as their
discussion is very similar.\ All processes associated with $\Sigma$, which
include those by D, occur \emph{internally} inside it. Either they refer to
flows in both directions between $\Sigma_{1}$ and $\Sigma_{2}$, or refer to
internal processes within them. Including D within $\Sigma$ allows us to avoid
the issue of separate work done by D; we only deal with the work done by
$\Sigma$. To emphasize the internal nature, we will use $d_{\text{i}}\varphi$
\cite{Prigogine} to denote changes caused by processes within $\Sigma
_{\text{b}}$. Maxwell has \emph{conjectured }that D causes a temperature
difference $\Delta T=T_{1}-T_{2}>0$ in the chambers. This is precisely how the
situation is commonly treated. This assumes that the temperature is well
defined even in NEQ states, which is not so obvious. No number imbalance is
treated separately. Being in a NEQ state, $\Sigma_{\text{b}}$ is spatially
nonuniform so identifying its \emph{unique} global (over entire $\Sigma
_{\text{b}}$) temperature $T_{\text{b}}=(T,T_{1},T_{2})$ as was first
envisioned by Planck \cite{Planck} is nontrivial. This requires identifying
the NEQ entropy for any arbitrary NEQ state
\cite{Gujrati-Entropy1,Gujrati-Entropy2}. Once the entropy is identified,
$T_{\text{b}}$ can be uniquely defined \emph{thermodynamically} for any NEQ
state of $\Sigma_{\text{b}}$ even if it is inhomogeneous as recently shown
\cite{Gujrati-I,Gujrati-II,Gujrati-III} and reviewed in \cite{Gujrati-Review},
where it is demonstrated that it satisfies all the sensible conditions
including the Clausius statement that are required of a global thermodynamic
temperature. Using this definition, which is explained later, we provide a
first-ever resolution of the second law paradox created by D that $\Delta
T\equiv0$ by using only classical thermodynamics. None of the resolutions
available so far to salvage the second law is based solely on thermodynamics.

We take a different approach from all previous approaches by recalling the
Clausius statement according to which $\Delta T\neq0$ generates
\emph{spontaneous} heat flow from hot to cold in accordance with the second
law and argue that it competes with Maxwell's conjecture. The temperature
difference $\Delta T$\ acts as a thermodynamic force \cite{Prigogine} that
brings back EQ so that $\Delta T=0$. This spontaneous heat flow is in response
to the destruction of EQ that D has attempted. Indeed, we provide a
generalization of the Clausius statement to other kinds of flow such as a mass
flow, \textit{i.e.}, other thermodynamic forces. By assuming $\Delta T\neq0$,
we shows that D remains subservient to thermodynamic forces at all times so it
\emph{never} succeeds in destroying EQ, which disproves Maxwell's conjecture.
We find that D unwittingly brings EQ back and ensures $\Delta T=0$ so he is
incapable of destroying EQ. This conclusion is similar to that by Smoluchowski
\cite{Smoluchowski} and Feynman \cite{Feynman}; only the reasoning is
different and is based solely on classical thermodynamics without fluctuations.

The rest of the paper is to provide a theoretical support for the above
scheme, using our recently developed nonequilibrium thermodynamics denoted by
MNEQT \cite{Gujrati-Review}. As it is a generalization of classical
thermodynamics \cite{deGroot,Kestin,Maugin,Prigogine}\ developed by Carnot,
Kelvin, Clausius, and Maxwell \cite{Prigogine}, our demonstration is as valid
as classical thermodynamics. The novelty of the MNEQT is that it is based on
system-intrinsic quantities that also include internal variables
\cite{deGroot,Kestin,Maugin,Prigogine} that \emph{uniquely} specify a NEQ
state of $\Sigma_{\text{b}}$ and its unique entropy
\cite{Gujrati-Entropy1,Gujrati-Entropy2} in $\mathfrak{S}_{\mathbf{Z}}$ so it
fully captures whatever is going on within $\Sigma_{\text{b}}$
\cite{Gujrati-Review} through $d_{\text{i}}\varphi$, which is what we are
interested in as discussed above. If a particular $d_{\text{i}}\varphi$ is
allowed in a process, its sign is controlled by the second law
\cite{Gujrati-Review,Prigogine} as proven below in Eq. (\ref{diS}).

In the MNEQT \cite{Gujrati-I,Gujrati-II,Gujrati-III,Gujrati-Heat-Work}, we
handle irreversibility directly using internal quantities $d_{\text{i}}%
S\geq0,d_{\text{i}}W\equiv d_{\text{i}}Q\geq0$, where $d_{\text{i}}S$ is the
irreversible entropy generation, and $d_{\text{i}}W$, $d_{\text{i}}Q$ are
irreversible work and heat generated within $\Sigma_{\text{b}}$; in contrast,
$d_{\text{i}}E=d_{\text{i}}Q-d_{\text{i}}W\equiv0$ as no internal process can
change the energy of the system, and $d_{\text{i}}N=0$ (no chemical reaction).
In classical thermodynamics, where $d_{\text{i}}W$, $d_{\text{i}}Q$ are not
recognized, irreversibility is assessed indirectly as an inequality
\cite{Landau,Prigogine,deGroot} by using $d_{\text{i}}S\geq0$. But, in the
MNEQT, we only deal with equalities, which makes our demonstration possible.

Let us briefly review the MNEQT for an arbitrary isolated system $\Sigma$,
composed of several quasi-independent and disjoint subsystems $\Sigma
_{j},j=1,2,\cdots,n$ so\ that the entropies of $\Sigma_{j}$ are additive to
give the entropy $S$ of $\Sigma$.\ We take $\Sigma$ to have fixed $N$, and
$E$. We will also assume that the volume of $\Sigma_{j}$ is held fixed at
$V_{j}=V/n$ but its energy $E_{j}$ and the number of particles $N_{j}$ can
vary but always satisfy $%
%TCIMACRO{\tsum \nolimits_{j}}%
%BeginExpansion
{\textstyle\sum\nolimits_{j}}
%EndExpansion
N_{j}=N,%
%TCIMACRO{\tsum \nolimits_{j}}%
%BeginExpansion
{\textstyle\sum\nolimits_{j}}
%EndExpansion
E_{j}=E$. We set $\mathbf{X}_{j}=(N_{j},E_{j})$, and as before use
$\Sigma_{\text{b}}$ for any of $\Sigma,\left\{  \Sigma_{j}\right\}  $.

In EQ, the entropy of $\Sigma_{\text{b}}$\ is a state function
$(S(E,N),\left\{  S_{j}(N_{j},E_{j})\right\}  )$ in $\mathfrak{S}_{\mathbf{X}%
}$. For $\Sigma$ in EQ, we have $N_{j}=N/n,E_{j}=E/n$. This is not the case
when $\Sigma_{\text{b}}$ is in a NEQ state
\cite{Gujrati-Entropy1,Gujrati-Entropy2}. It is useful to think of the two
distinct \emph{realizations} for $\Sigma_{\text{b}}$. We can treat it as a
"composite" body $\Sigma_{\text{C}}$ with \emph{detailed} information of its
subsystems. This realization is useful to explicitly consider processes such
as flows between its subsystems. Alternatively, we can treat $\Sigma
_{\text{b}}$ as a "black box" $\Sigma_{\text{B}}$ if we only need to
investigate its thermodynamics without any detailed information of internal
processes. As we show, these internal processes are described by internal variables.

For the moment, let us assume that each $\Sigma_{j}$ is in EQ with its entropy
$S_{j}(\mathbf{X}_{j}(t))$. The entropy $S_{\text{C}}$ of $\Sigma_{\text{C}}$
\begin{equation}
S_{\text{C}}(\left\{  \mathbf{X}_{j}(t)\right\}  )=%
%TCIMACRO{\tsum \nolimits_{j}}%
%BeginExpansion
{\textstyle\sum\nolimits_{j}}
%EndExpansion
S_{j}(\mathbf{X}_{j}(t))\leq S(\mathbf{X}(t)), \label{Entropy-C}%
\end{equation}
is a function of $2n$ independent variables for a state that is represented as
an $n$-tuple in $\mathfrak{S}_{\mathbf{X}}$. For $\Sigma_{\text{B}}$, we need
to use $N$ and $E$ as its observables to specify its entropy $S_{\text{B}}$.
From entropy additivity, it is given exactly by the right side of the first
equation in Eq. (\ref{Entropy-C}) so we need additional $2(n-1)$ independent
independent variables, the internal variables, shown collectively by
$\boldsymbol{\xi}$ to specify $S_{\text{B}}$ uniquely in $\mathfrak{S}%
_{\mathbf{Z}}$. For example, for $n=2$, the internal variables%
\begin{equation}
\xi_{\text{N}}=N_{1}(t)-N_{2}(t),\xi_{\text{E}}=E_{1}(t)-E_{2}(t)
\label{InternalVar-2}%
\end{equation}
constructed form $\left\{  N_{j}(t)\right\}  $ and $\left\{  E_{j}(t)\right\}
$ are independent of $N$ and $E$; see \cite{Gujrati-Review} for a general
discussion. The entropy $S_{\text{B}}$ of $\Sigma_{\text{B}}$ is a state
function and \emph{obeys} the second law
\begin{equation}
S_{\text{B}}(\mathbf{Z}(t))=%
%TCIMACRO{\tsum \nolimits_{j}}%
%BeginExpansion
{\textstyle\sum\nolimits_{j}}
%EndExpansion
S_{j}(\mathbf{X}_{j}(t))\leq S(\mathbf{X}(t)). \label{Entropy-B}%
\end{equation}
This state is called an \emph{internal EQ} (IEQ) state
\cite{Gujrati-I,Gujrati-II,Gujrati-III,Gujrati-Heat-Work} as it has the
maximum possible value for given $\mathbf{Z}(t)$. We also note the equality
$S(\mathbf{Z})\equiv S_{\text{B}}(\mathbf{Z})\equiv S_{\text{C}}(\left\{
\mathbf{X}_{j}\right\}  )$, with $S(\mathbf{Z})$ \emph{uniquely} defined for
an IEQ state $\mathcal{M}$ as each $S_{j}(\mathbf{X}_{j})$\ is uniquely
defined. It is trivial to generalize the above discussion to $\Sigma_{j}$'s in
IEQ with entropies $S_{j}(\mathbf{Z}_{j}(t))$ by treating each $\Sigma_{j}$
consisting of some subsubsystems.

We will assume \cite{Gujrati-Review,Gujrati-II} that any arbitrary NEQ state
can be described as an IEQ state by properly choosing an appropriate number of
internal variables. It happens that different $\xi_{l}\in\boldsymbol{\xi}$
have different relaxation times $\tau_{l}$ past which they equilibrate and do
not affect thermodynamics; see for example \cite{Gujrati-Hierarchy}. Thus, for
a given observational time scale $\tau_{\text{obs}}$, the time to make
consecutive measurements in an experimental setup, only those $\xi_{l}$'s need
to be considered for which $\tau_{l}>\tau_{\text{obs}}$. Thus, in practice,
the number of internal variables will be much smaller than $2(n-1)$. For
simplicity of discussion, we will consider only two internal variables shown
in Eq. (\ref{InternalVar-2}) for any body $\Sigma_{\text{b}}$\ as our aim is
only to demonstrate how classical NEQ thermodynamics can be used to show the
subservient nature of the demon. Adding more internal variables is not going
to affect the final conclusion.

In the following, we will also use $\beta^{\prime}$s to denote inverse
temperatures. The temperature and chemical potential of $\Sigma_{j}$ are given
by their standard EQ definition \cite{Landau,Prigogine}%
\begin{equation}
\beta_{j}=1/T_{j}=\partial S_{j}(\mathbf{X}_{j})/\partial E_{j},\beta_{j}%
\mu_{j}=-\partial S_{j}(\mathbf{X}_{j})/\partial N_{j}\label{Eq-Fields}%
\end{equation}
that appear in the EQ Gibbs fundamental relation
\begin{equation}
dS_{j}=\beta_{j}(dE_{j}-\mu_{j}dN_{j})\label{dS-EQ}%
\end{equation}
in $\mathfrak{S}_{\mathbf{X}}$, from which we can construct $dS_{\text{C}}$
using Eq. (\ref{Entropy-C})%
\begin{equation}
dS_{\text{C}}=%
%TCIMACRO{\tsum \nolimits_{j}}%
%BeginExpansion
{\textstyle\sum\nolimits_{j}}
%EndExpansion
\beta_{j}(dE_{j}-\mu_{j}dN_{j}).\label{dS-C}%
\end{equation}
The NEQ Gibbs fundamental relation \cite{Gujrati-Review,Gujrati-II}
\begin{equation}
dS_{\text{B}}=\beta(dE-\mu dN+\mathbf{A}\cdot d\boldsymbol{\xi})\label{dS-B}%
\end{equation}
for $S_{\text{B}}$ in $\mathfrak{S}_{\mathbf{Z}}$ leads to the Clausius
statement and yields
\begin{equation}
\beta=1/T=\partial S_{\text{B}}/\partial E,\beta\mu=-\partial S_{\text{B}%
}/\partial N,\beta\mathbf{A}=\partial S_{\text{B}}/\partial\boldsymbol{\xi
;}\label{NEQ-fields}%
\end{equation}
here, $\mathbf{A}$ is the affinity or thermodynamic force \cite{Prigogine}
associated with $\boldsymbol{\xi}$. As $\xi_{l}\in\boldsymbol{\xi}$
equilibrates, $A_{l}\in\mathbf{A}\rightarrow0$ and plays no role in a NEQ
process. Reexpressing Eq. (\ref{dS-B})\ as
\[
dE=TdS_{\text{B}}+\mu dN-\mathbf{A}\cdot d\boldsymbol{\xi,}%
\]
allows us to identify the generalized heat $dQ_{\text{B}}=TdS_{\text{B}}$ and
generalized work $dW_{\text{B}}=-\mu dN+\mathbf{A}\cdot d\boldsymbol{\xi}$. As
$\mathbf{A}\cdot d\boldsymbol{\xi}$ is generated by internal processes, we
must have $A_{l}d\xi_{l}\geq0$ in accordance with the second law
\cite{Gujrati-Review,Gujrati-II}; see Eq. (\ref{diS}) for proof. Each
$A_{l}d\xi_{l}\geq0$ is called the \emph{generalized Clausius statement} for
the internal variable or flow $\xi_{l}$.

For simplicity, take $n=2$, see Eq. (\ref{InternalVar-2}); the generalization
to any $n$ is trivial. Then, $\mathbf{A}=\left(  A_{\text{N}},A_{\text{E}%
}\right)  $ and
\begin{equation}
dS_{\text{B}}=\beta\lbrack dE-\mu dN+A_{\text{E}}d\xi_{\text{E}}+A_{\text{N}%
}d\xi_{\text{N}}]. \label{dS-B2}%
\end{equation}
Equating it with $dS_{\text{C}}$, we easily establish \cite[see 10.3.1]%
{Gujrati-Review}
\begin{subequations}
\label{T-A-2}%
\begin{align}
\beta &  =\frac{\beta_{1}+\beta_{2}}{2},\beta\mu=\frac{\beta_{1}\mu_{1}%
+\beta_{2}\mu_{2}}{2},\label{Fields-2}\\
\beta A_{\text{E}}  &  =\frac{\beta_{1}-\beta_{2}}{2},\beta A_{\text{N}}%
=\frac{\beta_{1}\mu_{1}-\beta_{2}\mu_{2}}{2}, \label{affinities-2}%
\end{align}
which expresses $T$ and $\mu$ of $\Sigma_{\text{B}}$ in terms of those of
$\Sigma_{1}$ and $\Sigma_{2}$ in $\Sigma_{\text{C}}$. In EQ $\Sigma$,
$\beta_{1}=\beta_{2}$ and $\mu_{1}=\mu_{2}$ as expected.

We should emphasize that taking $\Sigma_{\text{B}}$ to consist of more than
two subsystems will require additional internal variables for $S_{\text{B}}$,
but the lesson here is that in all cases, we can identify a unique global NEQ
temperature, chemical potential, and affinities for the realization
$\Sigma_{\text{B}}$. We do not need to know its internal structure explicitly,
which is fully captured by internal variables.

To understand the physics of $A_{\text{E}}d\xi_{\text{E}}$ and $A_{\text{N}%
}d\xi_{\text{N}}$ for internal processes ($d_{\text{i}}N=d_{\text{i}}E=0$) in
$\Sigma_{\text{b}}$, we fix $N$ and $E$. Equating Eq. (\ref{dS-B2}) with
$d_{\text{i}}S_{\text{C}}=\left[  \beta_{1}-\beta_{2}\right]  dE_{1}%
(t)+\left[  \beta_{1}\mu_{1}-\beta_{2}\mu_{2}\right]  dN_{1}(t)\geq0$ from Eq.
(\ref{dS-C}) gives
\end{subequations}
\begin{subequations}
\label{diS}%
\begin{align}
d_{\text{i}}S^{\text{Q}}  &  =\left[  \beta_{1}-\beta_{2}\right]
dE_{1}(t)=\beta A_{\text{E}}d\xi_{\text{E}}(t)\geq0,\label{dis-Q}\\
d_{\text{i}}S^{\text{N}}  &  =\left[  \beta_{1}\mu_{1}-\beta_{2}\mu
_{2}\right]  dN_{1}(t)=\beta A_{\text{N}}d\xi_{\text{N}}(t)\geq0,
\label{dis-N}%
\end{align}
for the energy and particle flows in $\Sigma_{\text{b}}$ at fixed $\mathbf{X}%
$. From the general expression for $dW_{\text{B}}$ and $dQ_{\text{B}}$ above,
we also have $d_{\text{i}}W_{\text{B}}=\mathbf{A}\cdot d\boldsymbol{\xi
}=Td_{\text{i}}S=T(d_{\text{i}}S^{\text{Q}}+d_{\text{i}}S^{\text{N}})\geq0$
for any number of internal variables in the isolated $\Sigma_{\text{b}}$.

After identifying the NEQ $T$ and $\mu$ for $\Sigma_{\text{b}}$ in the MNEQT,
we turn to the demon problem in which D, according to Maxwell's conjecture,
causes $\Sigma$ to leave $\mathcal{M}_{\text{eq}}\in\mathfrak{S}_{\mathbf{X}}%
$\ at $t=0$ to go into $\mathcal{M}\in\mathfrak{S}_{\mathbf{Z}}$ at $t>0$ due
to each of the two flows in both directions across the wall and the hole.
Following his conjecture, we assume $\Delta T>0$, and nonzero $\mathbf{\xi
}(t)$ with $\mathbf{\xi}_{\text{eq}}=0$. Then
\end{subequations}
\begin{equation}
\Delta S_{\text{D}}\doteq S_{\text{B}}(\mathbf{X,\xi}(t))-S_{\text{B}%
}(\mathbf{X})<0 \label{Entropy-Change-Demon}%
\end{equation}
is the entropy loss of $\Sigma$ caused by the demon, a seeming violation of
the second law.

We now show that this is not the complete story as this loss is
counterbalanced by a spontaneous equilibration process specified by
$S_{\text{B}}(\mathbf{X,\xi})$ for $\mathcal{M}$\ in accordance with the
generalized Clausius statement, see Eq. (\ref{diS}), on which D has no
control, and forces $\mathcal{M}\rightarrow\mathcal{M}_{\text{eq}}$.  The
spontaneous process is integral to thermodynamic consideration so it must not
be neglected as the case has been so far. We take $\Sigma_{\text{b}}$ to refer
to the chamber $\Sigma_{j},j=1,2$, which we treat as a NEQ $\Sigma_{\text{B}}$
and use Eq. (\ref{NEQ-fields}) to identify its $\beta_{j}$, $\mu_{j}$, and
$\mathbf{A}_{j}$ by treating $\Sigma_{j}$ to consists of two EQ subchambers
$\Sigma_{j1}$ and $\Sigma_{j2}$. This implies $n=4$ subchambers for $\Sigma$.
The interface between $\Sigma_{21}$ and $\Sigma_{12}$ is the above wall in
$\Sigma$. We have restricted to only two internal variables for $\Sigma_{j}$
as said above, but more internal variables can be taken for a more complex NEQ
chamber. The $(\mathbf{A}_{j},\boldsymbol{\xi}_{j})$ only refer to spontaneous
processes between $\Sigma_{j1}$ and $\Sigma_{j2}$\ that partly drive
$\mathcal{M}\rightarrow\mathcal{M}_{\text{eq}}$, for which at fixed
$\mathbf{X}_{j}$, we have $d_{\text{i}}S_{j}^{\text{Q}}=\beta_{j}A_{\text{E}%
j}d\xi_{\text{E}j}(t)\geq0,d_{\text{i}}S_{j}^{\text{N}}=\beta_{j}A_{\text{N}%
j}d\xi_{\text{N}j}(t)\geq0$ from Eq. (\ref{diS}), and $d_{\text{i}}%
W_{\text{B}j}=T_{j}d_{\text{i}}S_{j}=T_{j}(d_{\text{i}}S_{j}^{\text{Q}%
}+d_{\text{i}}S_{j}^{\text{N}})\geq0$. These internal processes in $\Sigma
_{j}$, which have never been discussed before, do not depend on the nature of
the wall.

The complete equilibration $\mathcal{M}\rightarrow\mathcal{M}_{\text{eq}}$
requires additional flows, never discussed in the literature, between
$\Sigma_{21}$ and $\Sigma_{12}$ that change $\mathbf{X}_{j}$. This is where
the nature of the wall becomes important. However, it is much simple to treat
$\Sigma$ as $\Sigma_{\text{C}}$ having the two NEQ chambers $\left\{
\Sigma_{j}\right\}  $, each treated as a NEQ $\Sigma_{\text{B}}$ with its own
$T_{j},\mu_{j},A_{\text{E}j}$, and $A_{\text{N}j}$ without knowing its
interior. 

We first deal with the permeable wall. The flows occur across the wall and the
open hole. As $\Sigma$ has $n=4$ subchambers, we need $6$ internal variables
in Eq. (\ref{Entropy-Change-Demon}) to uniquely specify $\mathcal{M}$\ for
$\Sigma$, each one equilibrating with its own relaxation time. To simplify the
discussion, we again restrict to only two internal variables; see Eq.
(\ref{InternalVar-2}). Thus, $d_{\text{i}}S_{\text{B}}^{\text{Q}}=\left[
\beta_{1}-\beta_{2}\right]  dE_{1}(t)\geq0,d_{\text{i}}S_{\text{B}}^{\text{N}%
}=\left[  \beta_{1}\mu_{1}-\beta_{2}\mu_{2}\right]  dN_{1}(t)\geq0$ and
$d_{\text{i}}W_{\text{B}}=Td_{\text{i}}S_{\text{B}}=T_{j}(d_{\text{i}%
}S_{\text{B}}^{\text{Q}}+d_{\text{i}}S_{\text{B}}^{\text{N}})\geq0$ due to
these flows at fixed $\mathbf{X}$ across the wall, see Eq. (\ref{diS}), as
$\mathcal{M}\rightarrow\mathcal{M}_{\text{eq}}$.

We note that $dE_{1}(t)$\ and $dN_{1}(t)$ each have three independent
contributions: from the wall, from the particles controlled by D, and from the
additional flows through the open hole. In general, $\mathbf{\xi}$ has six
internal variables, which we now assume have been included in the
determination of $d_{\text{i}}S_{\text{B}}$. After equilibration,
$S(\mathbf{Z})$ increases by
\begin{equation}
\Delta S_{\text{CS}}\doteq S_{\text{B}}(\mathbf{X})-S_{\text{B}}%
(\mathbf{X,\xi}(t))>0 \label{Entropy-Change-Clausius}%
\end{equation}
in accordance with the generalized Clausius statement. We see that $\Delta
S\doteq\Delta S_{\text{D}}+\Delta S_{\text{CS}}=0$, thus satisfying the second
law. The situation becomes more clear by considering an infinitesimal process
in $\mathcal{M}_{\text{eq}}$ for which
\begin{equation}
d_{\text{i}}S_{\text{B}}\equiv dS\doteq dS_{\text{D}}+dS_{\text{CS}}=0
\label{Entropy-Conservation}%
\end{equation}
that shows that $\mathcal{M}_{\text{eq}}$ never leaves itself due to the
infinitesimal spontaneous counterbalancing process after D attempts to
destroys it: the loss $dS_{\text{D}}$ is immediately recovered by irreversible
entropy generation $dS_{\text{CS}}$. In essence, no temperature or entropy
difference ever arises so that $\Delta T=0\Rightarrow T_{\text{b}}%
=T_{0}\Rightarrow dS_{\text{D}}=dS_{\text{CS}}=0$, and $d_{\text{i}}W\equiv
d_{\text{i}}W_{\text{B}}=0$ at all times.

Let us now consider the two cases of an impervious wall. We will argue that
nothing changes in the discussion above except for the definitions of
$dE_{1}(t)$ and $dN_{1}(t)$. (a) There is no flow across the wall but flows of
all sorts of particles, including those controlled by D, are allowed through
the open hole determining $dE_{1}(t)$ and $dN_{1}(t)$ above. (b) Only the
flows caused by the particles that D controls through the open hole determine
$dE_{1}(t)$ and $dN_{1}(t)$.

There is an intuitive way to understand the physics in (b), which reveals some
surprising facts about the effect of D. Take $\Delta T>0$ following Maxwell's
conjecture. As D allows faster particles to add $dE_{1}^{\prime}>0$, not to be
confused with $dE_{1}$ for (b) above, into $\Sigma_{1}$, he \emph{decreases}
$S$ by $dE_{1}^{\prime}(\beta_{2}-\beta_{1})$. Surprisingly, as D allows
slower particles to add $dE_{2}^{\prime}=-dE_{1}^{\prime}$ into $\Sigma_{2}$,
$S$\ \emph{increases} by $dE_{2}^{\prime}(\beta_{1}-\beta_{2})$; cf. Eq.
(\ref{dis-Q}). As they cancel each other without changing $S$, Maxwell's
conjecture $\Delta T\neq0$ cannot be justified if we start in $\mathcal{M}%
_{\text{eq}}$: D never destroys EQ.

This is no different from what happens in $\mathcal{M}_{\text{eq}}$ with the
permeable wall, across which each flow cancels out when both directions are
considered as noted earlier with no change in the entropy. The same happens
for the flows of all sorts of particles through the open hole. Thus, unbeknown
to\ Maxwell, he actually had designed the demon to support the second law as
$\Sigma$ never leaves $\mathcal{M}_{\text{eq}}$, which is contrary to the
popular belief in physics and philosophy. Maxwell's conjecture is false in all cases.

It is gratifying to see after all that no internal process can drive an
isolated system away from EQ. The MNEQT is used to identify global $T$ and
$\mu$, and internal variables to test Maxwell's conjecture. The importance of
our thermodynamic $T$ and $\mu$\ is that they satisfies Clausius's general
statement in any $\mathcal{M}$ as proved in Eq. (\ref{diS}). Without such a
proof, our demonstration will just be another conjecture. As thermodynamics is
devoid of fluctuations, our reasoning is somewhat different from those offered
by Smoluchowski \cite{Smoluchowski} and by Feynman \cite{Feynman}. Recently,
Hoover and Hoover \cite{HHoover} have carried out a two-dimensional simulation
of particles with interaction to test our conclusion. They find that the heat
conductivity in the model competes with the demon's goal as we have concluded. 

The discussion also reveals something very profound. If the demon only allows
the particles to move in only one direction, then such a demon will actually
violate the second law according to our analysis. It will be interesting to
pursue if attempts involving information and erasure, etc. can salvage the
second law for such a demon.

Comments from S. Ciliberto, D. Kondepudi, and J.D. Norton are gratefully
acknowledged. We also thank Bill Hoover and Carol Hoover for sharing their result.

\end{document}